\documentstyle[epsfig,psfig,12pt]{report}
\def\deg{\hbox{$^\circ$}}

\def\arcsec{\hbox{$^{\prime\prime}$}}
\def\fmag{\hbox{$.\!\!^m$}}

\begin{document}

\vspace{3cm}
\centerline{\Huge{\bf{The FIRST-APM QSOs Survey}}}
\vspace{0.5cm}
\centerline{\Huge{\bf{in the SBS sky region\footnotemark{}}}}

\footnotetext{Based on observations from the Guillermo Haro
Astrophysical Observatory, Cananea, M\'exico} 

\vspace{1cm}

\centerline{\it 3rd Guillermo Haro Astrophysics Conference on} 

\centerline{\bf Cosmic Evolution and Galaxy Formation} 

\centerline{\it Puebla, Mexico, November 15-19, 1999} 

\vspace{0.5cm}

\vspace{2cm}

\centerline{\Large \bf Vahram H. Chavushyan}
\vspace{0.5cm}
\centerline{\it Instituto Nacional de Astrof\'{\i}sica Optica y Electr\'onica.}
\centerline{\it Apartado Postal 51 y 216. C.P. 72000. Puebla, Pue., M\'exico}
\centerline{\it e-mail: vahram@inaoep.mx}
\vspace{1cm}

\centerline{\Large \bf Oleg V. Verkhodanov}
\vspace{0.5cm}
\centerline{\it Special Astrophysical Observatory RAS. Nizhnij Arkhys,}
\centerline{\it Karachai-Cherkesia, 357147. Russia.}
\centerline{\it email: vo@sao.ru}

\vspace{1cm}

\centerline{\Large \bf Ra\'ul M\'ujica and Luis Carrasco}
\vspace{0.5cm}
\centerline{\it Instituto Nacional de Astrof\'{\i}sica Optica y Electr\'onica}
\centerline{\it Apartado Postal 51 y 216. C.P. 72000. Puebla, Pue., M\'exico}
\centerline{\it e-mail: rmujica@inaoep.mx, carrasco@inaoep.mx}

\newpage

\centerline{\bf \Large ABSTRACT}

\vspace{0.5cm}

\large 
  
The main goal of the FIRST-APM QSO Survey (FAQS) survey is to compile
the most complete
sample of Bright QSOs, located in the well optically investigated area of the sky covered
by the Second Byurakan Survey (SBS). We do that through the combination of
both radio and optical surveys, down to the magnitude limit $B\le18\fmag5$.
We report here the first results of our ongoing study, that is based upon
the cross-identification of the FIRST radio catalog and the Automated
Plate Measuring Machine (APM) optical catalog. The overlapping sky area 
between the FIRST and the SBS surveys is
about 700 deg$^{2}$. Our compiled list of sources for this overlapping
region contains $\sim 400$ quasar candidates brighter than $B=18\fmag 5$.
Of which, about 90 objects are already spectroscopically classified. These objects have been discovered maynly by the SBS survey. During 1999,
we have carried out spectroscopic observations for more than 100 FAQS
objects with the 2.1m telescope of the Guillermo Haro
Astrophysical Observatory (GHO). So far, in the studied subsample,
we have found 33 new QSOs, 2 Seyfert Galaxies, 15 emission line galaxies,
1 BL Lac, and 45 high galactic latitude stars. Amongst the 33 QSOs, we have found
two broad absorption
line (BAL) QSOs, namely, FAQS 151113.7+490557 and FAQS 161744.4+515054.
These two BAL QSOs are radio-loud, and have radio-to-optical flux ratios
({\it log R}) with values of 1.1 and 2.5 respectively. The last object being
the brightest radio-loud BAL QSO known, with a flux density of 99.8 mJy
at 1.4~Ghz.

\vspace{0.5cm}

\centerline{\bf \Large INTRODUCTION}

\vspace{0.5cm}

QSO surveys may provide basic information to understand a number of cosmological key issues, such as:
the epoch of galaxy formation,
the physics of the AGN phenomenon,
the formation and evolution of large scale structure.
So far, over 11000 QSOs are known (V\'eron-Cetty and
V\'eron 1998).
Several fundamental questions are still open in relation to the nature
of these objects and the internal relationships between their physical
properties. For instance, their surface and space densities are quite
important for the understanding of evolutionary aspects.
Yet, the latter requires always data for complete samples of objects,
both in terms of redshift and in terms of intrinsic luminosity.

The ideal searching technique should produce a sample that includes 100\%
of the objects of interest and
that is free of contaminating objects. However, in practice, one faces
a trade-off between
maximizing the level of completeness of the sample,
while at the same time, minimizing the number of confusing objects within it.
The degree of completeness can in theory be determined by comparing
the results from different survey techniques (e.g. multicolor selection
vs slitless spectroscopy).

Historically, the optical QSO samples have been selected either
on the basis of their emission
lines or peculiar colors. The QSOs can also be found by their radio,
X-ray and/or their infrared properties and/or their variable nature.
Optically selected quasar samples are the most complete ones. However, every survey technique has a redshift
and luminosity-dependent selection biases
(Wampler and Ponz 1985).
Taking into account the selection effects,
inherent to every technique adopted to search for QSOs, is obvious that only
the combination of the different search techniques for the different
spectral ranges, will yield a complete sample of quasars,
with an adequate representation of all the properties inherent
to them (Hartwick and Shade 1990, Chavushyan 1995).

The emphasis on optically selected samples is due to the efficiency of the
techniques in finding the objects of interest.
Radio selected samples suffer the difficulty of identifying
their optical counterparts, a difficulty arising from poor radio positions.
Furthermore, in the past, radio selected samples have been largely
insensitive to radio quiet objects, which
make up the majority of the QSO population.
Yet, optically selected quasar samples have their own disadvantages.
These samples usually exclude QSOs with colors different than typical.
A radio selected sample that reaches faint
flux limits, such as FIRST, shall be immune to optical color selection
effects.

The FIRST radio survey (Becker et al. 1995) provides a new resource
for constructing a large quasar sample. With positions accurate
to better than $1\arcsec$ and a point source sensitivity limit of 1 mJy,
it goes 50 times fainter in flux, than any previous radio survey.
A QSO survey based on FIRST and the APM (Irwin 1998) catalog is the FBQS
(Gregg et al. 1996). Unfortunately, in the FBQS area there is a lack
of complete optically selected samples of QSOs. It is
hence, very difficult to estimate the completeness of the survey,
as well as, the number of missed quasars by the optical sample
conformed by the FBQS.

In order to circumvent this problem, we have started a multiwavelength
search of QSOs in the well investigated
Second Byurakan Survey (SBS) sky area, we also made use of radio and
X-ray data. The SBS is a low resolution objective prism survey with
a limiting magnitude of $B\sim19\fmag5$, that covers 1000 square degrees
in the sky region defined by $07^h 43^m < R.A. < 17^h 17^m$ and
$+49\deg < Dec.< +61\deg$ (Markarian and Stepanian 1983; Stepanian 1994a).
The SBS has produced one of the largest and most homogeneous complete
sample of bright QSOs (Stepanian 1994b; Stepanian et al. 1999) with accurate {\it BV} photometry (Chavushyan et al. 1995, 1999).
Amongst the bright quasar surveys the SBS is the only slitless survey
which covers a  large, continuous area of the sky. In this survey,
both stellar and non-stellar objects are selected jointly.

\vspace{0.5cm}

\centerline{\bf \Large SAMPLE}
\vspace{0.5cm}

\large

As a first step, in order to detect the missing QSOs in the SBS,
and to create the most complete QSO sample. We have used the deep radio
(FIRST) and optical (APM) databases, we call this sample FAQS.
The overlapping area between FIRST and SBS surveys is quite significant,
covering about 700 deg$^{2}$. We have cross-identified the FIRST catalog 
with APM objects. Adopting the following selection criteria:

\begin{enumerate}

\item Coincidence of the positions of the FIRST and APM objects within
a $3\arcsec$ radius; 

\item Objects classified as stellar-like on APM;

\item APM B--magnitudes between $14\fmag5$ and $18\fmag5$

\end{enumerate}

A procedure of cross-identification has been made with the CATS database
routines (Verkhodanov et al., 1997, {\it http://cats.sao.ru}), operating with incorporated
FIRST catalogue of the 1999 July version. \\ 

Our list of sources in this overlapping region contains $\sim 400$ objects,
of which 90 were previously known
AGNs (mainly discovered by the SBS).During 1999, we have carried out spectroscopic observations for about
100 FAQS objects. 

\vspace{0.5cm}

\centerline{\bf \Large OBSERVATIONS}
\vspace{0.5cm}

Three observing runs in 1999 were allocated to our project.
The observations were carried out with the 2.1m  GHO telescope and
the LFOSC (Zickgraf et al. 1997) focal reducer equipped with a $600\times400$ pixel CCD.
The read out noise of the detector being 8 $e^-$.
A set-up covering the spectral range of 4200-9000
\AA\  with a dispersion of 8 \AA /pixel was adopted.
The effective instrumental spectral resolution being about 16 \AA.\\

\vspace{0.5cm}

\centerline{\bf \Large PRELIMINARY RESULTS}
\vspace{0.5cm}

In this subsample we have found 33 new QSOs, 2 Seyfert Galaxies,
15 emission line galaxies, 1 BL Lac, and 45 high galactic latitude stars.\\

Among the 33 QSOs, we have found two broad absorption line (BAL) QSOs,
namely, FAQS 151113.7+490557 and
FAQS 161744.4+515054. These BAL QSOs are radio-loud,
and have a radio-to-optical
flux ratios, {\it log R} (Weymann et al. 1991) equal 1.1, and 2.5 respectively.
The last object has a flux density of 99.8 mJy at 1.4 GHz, being the brightest radio-loud BAL QSO known so far.\\

This long-term project will allow us to compile the first multiwavelength
complete sample of bright
quasars. Such a new complete sample of QSOs will help to address several
unsolved questions, such as
the surface density of bright quasars and whether or not there
is differential evolution between the radio-loud and the radio quiet QSOs.\\

This work has been supported by {\bf CONACYT} research
grants {\bf No. 28499-E},  and {\bf No. G2858-E}

\newpage

\vspace{0.5cm}

\centerline{\bf \Large REFERENCES}

\vspace{0.5cm}

\large

\begin{itemize}

\item[ ] Becker, R.H., White, R.L., and Helfand, D.J. 1995, ApJ, 450, 559

\item[ ] Chavushyan, V.H. 1995, PhD thesis. Nizhnij Arkhiz

\item[ ] Chavushyan, V.H., Stepanian, J.A., Balayan, S.K., and Vlasyuk, V.V.
1995, Astr. Lett. 21, 804

\item[ ] Chavushyan, V.H. et al. (in preparation)

\item[ ] Gregg, M.D., Becker, R.H., White, R.L., Helfand, D.J., McMahon,
R.G., and Hook, I.M. 1996, AJ, 112, 407

\item[ ] Hartwick, F.D.A., and Shade, D. 1990, ARAA, 28, 437

\item[ ] Irwin, M. 1998. {\it http://www.ast.cam.ac.uk/$\sim$apmcat/}

\item[ ] Markarian, B.E., and Stepanian, J.A. 1983, Astrofizika, 19, 639

\item[ ] Stepanian, J.A. 1994a, DSci thesis. Nizhnij Arkhiz

\item[ ] Stepanian, J.A. 1994b. In "Astronomy from Wide-Field Imaging". 
Eds. H.T. MacGillivray et al., 731 

\item[ ] Stepanian J.A., Chavushyan, V.H., Carrasco, L., Tovmassian, H.M.,
and Erastova, L.K. 1999, PASP, 111, 1099

\item[ ] Verkhodanov, O.V., Trushkin, S.A., Andernach, H., Chernenkov, V.N.
1997. In "Astronomical Data Analysis Software and Systems VI".
Eds. G. Hunt \& H.E. Payne. ASP Conference Series, 125, 322

\item[ ] V\'eron-Cetty, M.-P., and  V\'eron, P. 1998, A Catalog of Quasars
and Active Nuclei, ESO Sci. Rep. No. 18

\item[ ] Wampler, E.J., and Ponz, D. 1985, ApJ, 298, 448

\item[ ] Weymann, R.J., Morris, S.L., Foltz, C.B., and Hewett, P.C. 1991,
ApJ, 373, 23

\item[ ] Zickgraf, F.J., Thiering, I., Krautter, J., Appenzeller, I., Kneer,
R., Voges, W.H., et al. 1997, A\&AS, 123, 103

\end{itemize}

\newpage

%\vspace{2.5cm}

\begin{figure}
%[htb]
\epsfxsize=17cm\epsfbox{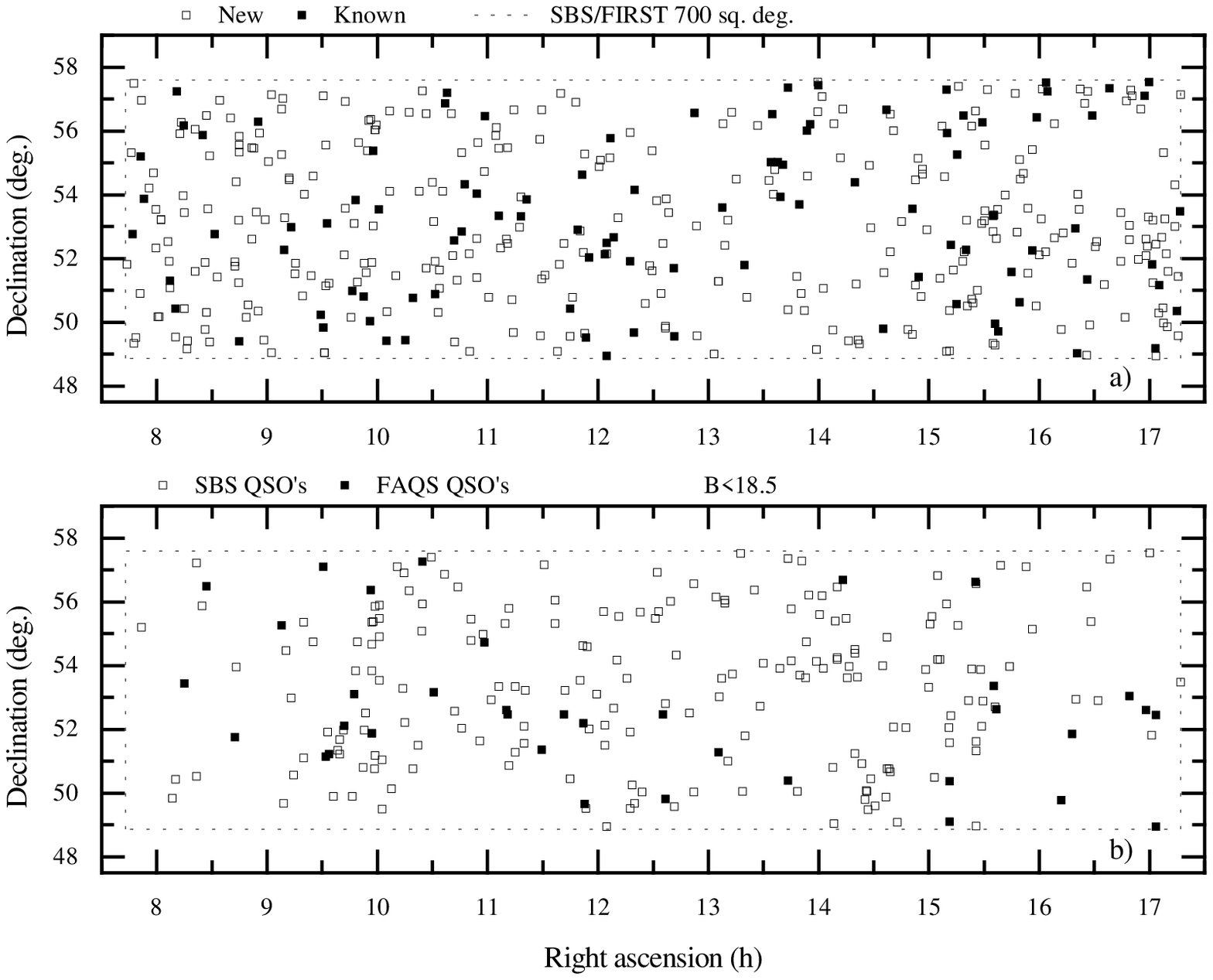}
\caption{a) The distribution of the 412 FAQS objects on the sky.
Open squares representing new QSO candidates and filled squares
representing the previously
known AGNs. b) The distribution of the SBS quasars (open squares)
and newly discovered FAQS QSOs (filled squares) on the sky.}
\end{figure}

\begin{figure}[htb]
\epsfxsize=17cm\epsfbox{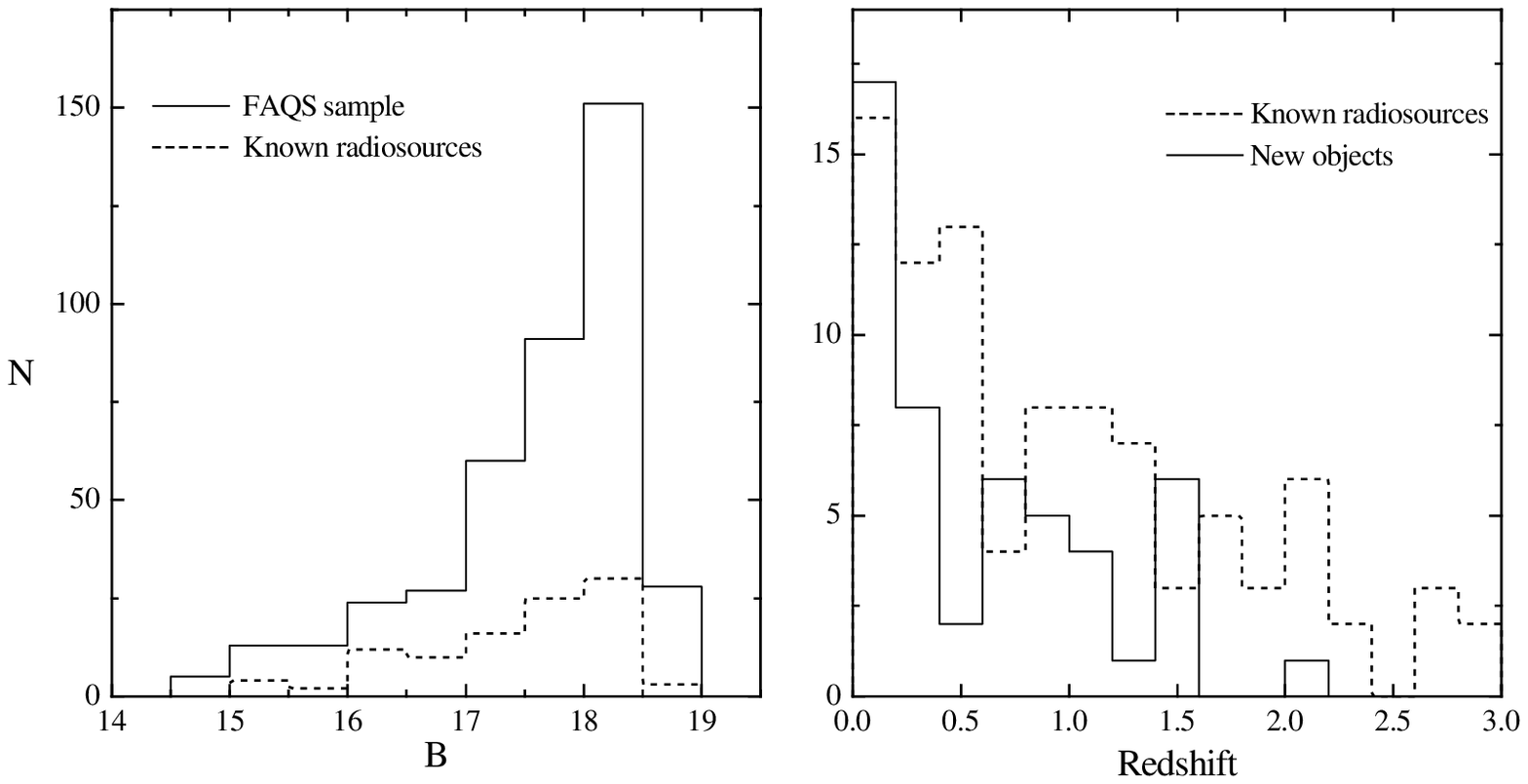}
\caption{The histograms of the magnitude and redshift
distributions for the FAQS sample.}
\end{figure}

\begin{figure}[htb]
\epsfxsize=17cm\epsfbox{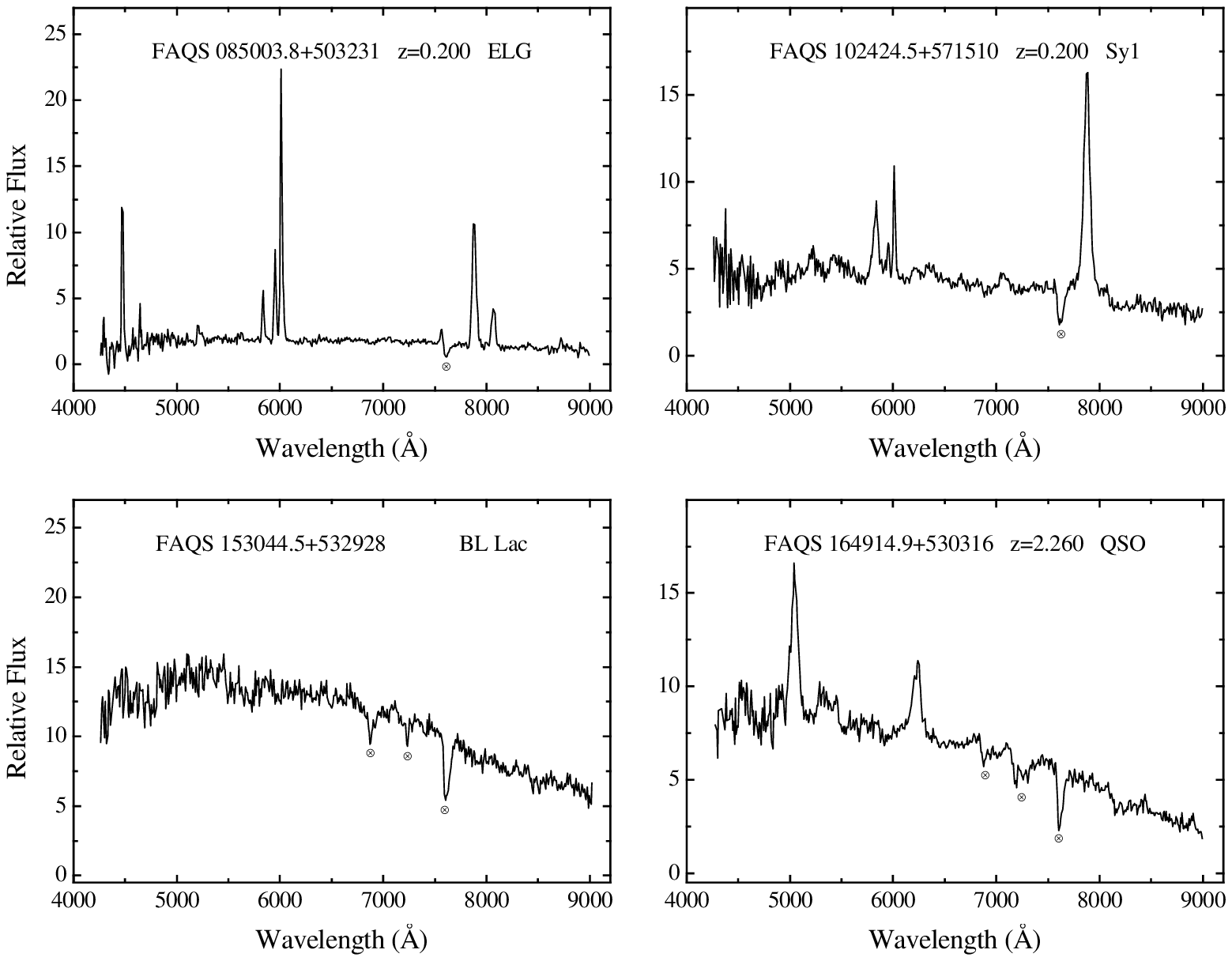}
\caption{Typical examples of LFOSC spectra of different types objects contained in the FAQS sample. 
There teluric absorption features have been marked ({\it encircled crosses}).}
\end{figure}

\begin{figure}[htb]
\epsfxsize=17cm\epsfbox{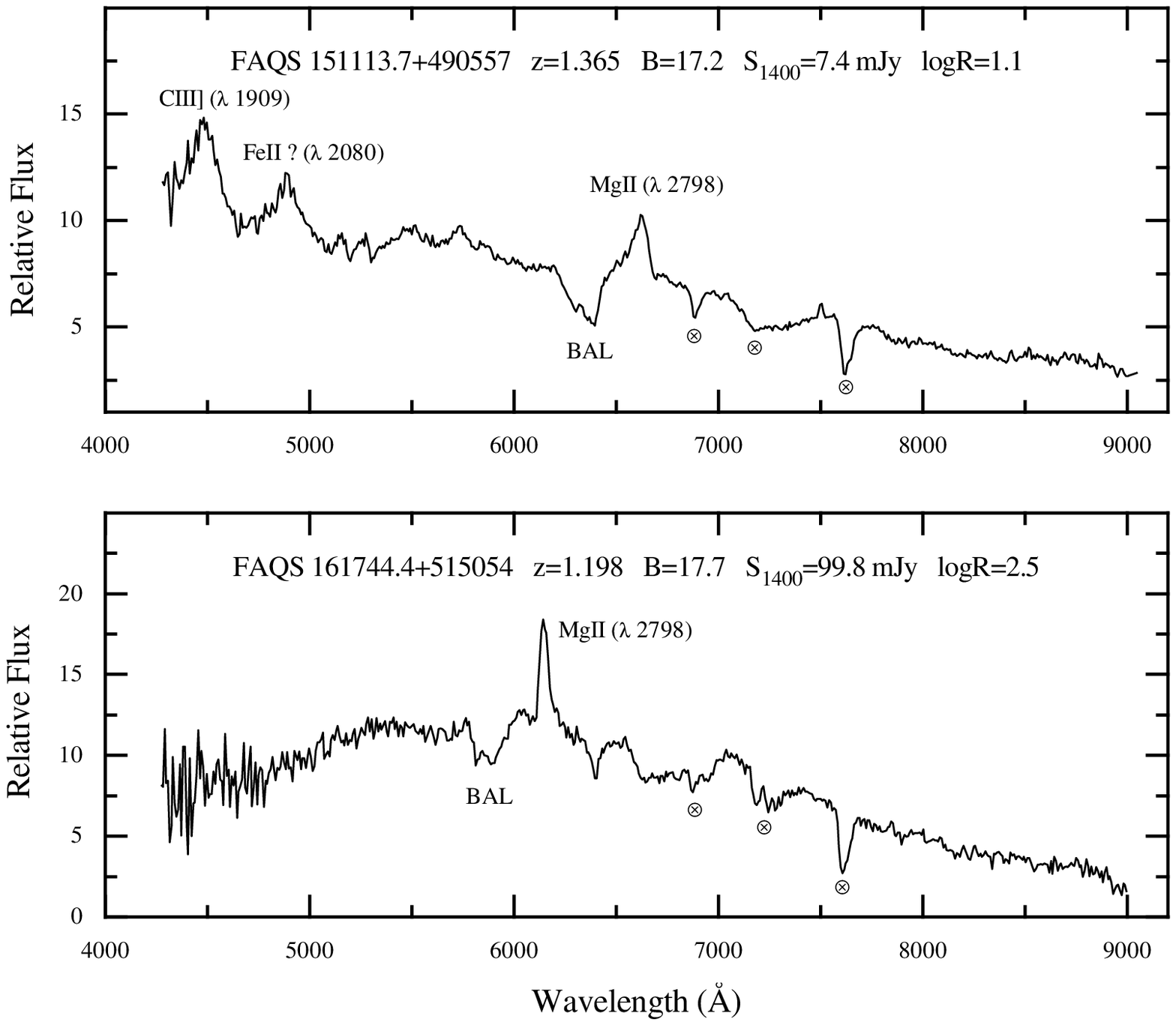}
\caption{The spectra of the radio-loud BAL QSOs found in the FAQS.
There teluric absorption features have been marked ({\it encircled crosses}).}
\end{figure}

\end{document}